\documentclass[12pt]{article}
\usepackage{newtxtext,newtxmath}
\usepackage{graphicx}
\usepackage[letterpaper,margin=1in]{geometry}
\renewenvironment{abstract}
	{\quotation}
	{\endquotation}

\date{}

\makeatletter
\renewcommand{\fnum@figure}{\textbf{Fig. \thefigure}}
\renewcommand{\fnum@table}{\textbf{Table \thetable}}
\makeatother

\usepackage[super,comma,sort&compress]{natbib}
\usepackage{doi} 
\usepackage{hyperref} 
\hypersetup{colorlinks=true, linkcolor=blue, citecolor=blue, urlcolor=blue}

\usepackage{soul}     
\usepackage{xcolor}    
\sethlcolor{yellow}

\soulregister\cite7 
\soulregister\ref7

\usepackage{url}
\usepackage{xurl}

\def\naturetitle{Electrostatic Depletion Force in Complex Coacervates}

\title{\bfseries \boldmath \naturetitle}

\author
{Zongpei Wu,$^{1}$ Shensheng Chen$^{1\ast}$\\
\normalsize{$^{1}$Department of Chemical and Biological Engineering,} \\
\normalsize{The Hong Kong University of Science and Technology,} \\
\normalsize{Clear Water Bay, Kowloon, Hong Kong, China}\\
\normalsize{$^\ast$Corresponding authors. E-mail: shensheng@ust.hk} \\
}

\date{}

\begin{document} 




\maketitle 

\begin{abstract} \bfseries \boldmath
The functionalities and applications of complex coacervates --- liquid condensates resulting from liquid-liquid phase separation of charged polymers --- are significantly influenced by the dispersion and aggregation states of guest macromolecules. Intriguingly, guest macromolecules exhibit a strong tendency to aggregate within coacervates even in the absence of apparent chemical incompatibility, indicating a universal aggregation mechanism at play in these environments. Using extensive MD simulations, we identify electrostatic depletion --- a strong force arising from electrostatic correlations within the host polyelectrolyte network that drives guest aggregations. Due to electrostatic depletion, neutral polymers, low-charge-density polyelectrolytes, and intrinsically disordered proteins (IDPs) exhibit effective attractions in coacervates, in stark contrast to their behavior in dilute solutions. Unlike traditional depletion effect that requires mismatched length scale and morphology, electrostatic depletion is relevant in fluid systems where solute and solvent are both polymers with comparable size. Our discovery bridges a critical knowledge gap in the molecular physics of densely charged, crowded liquids and holds significant implications for the design of synthetic protocells and advanced drug delivery systems. 
\end{abstract}

\section*{INTRODUCTION}
Life relies on the intricate functions of biomacromolecules operating within the crowded environments of living cells. Complex coacervates --- macromolecular condensates resulting from electrostatic associations of oppositely charged polymers such as proteins and RNAs --- are increasingly recognized as crucial mediums for intracellular activities\cite{banani2017biomolecular, deshpande2019spatiotemporal, abbas2021peptide, antifeeva2022liquid}. These charge-rich, liquid-like coacervates create microenvironments that selectively partition biomacromolecules and biochemical reactions, thereby regulating cellular organization and biochemistry. Inspired by their unique ability to encapsulate functional polymers, proteins, and enzymes in aqueous environments, researchers are  exploring coacervate-based systems for innovative applications in artificial cells  \cite{johnston2021associative, xu2022living, cook2023complex}, biomedicine \cite{black2014protein, blocher2020protein}, and functional nanoreactors \cite{schoonen2016compartmentalization, cao2023controlled}. Central to coacervates’ functionality, both in natural and synthetic contexts, is the aggregation/dispersion behavior of the encapsulated guest polymers, which governs intermolecular interactions, guest stability, and encapsulation efficiency. However, despite significant advances in characterizing coacervates themselves \cite{sing2020recent, rumyantsev2021polyelectrolyte}, the fundamental principles underlying guest polymer behavior --- particularly the mechanisms driving their aggregation and dispersion within the densely charged, crowded coacervate matrix --- remain poorly understood.        

Coacervates are multi-component polymer solutions, where polymer mixing/demixing behavior is long-predicted by the famous Flory-Huggins theory \cite{flory1942thermodynamics,huggins1942some}. The Flory-Huggins parameter $\chi$, arising from the differences in the short-range interactions among different species (chemical incompatibility), determines the degree of demixing. In the absence of any chemical incompatibility, ie., $\chi=0$ among all species, all polymer species are expected to be well-mixed with each other due to gained entropy upon mixing. This polymer physics paradigm has led to the predictions that for concentrated systems such as polymer blends \cite{knychala201750th} and coacervates \cite{Brangwynne2015PolymerTransitions}, relatively large $\chi$ between species is needed to induce demixing/aggregation due to the heavy screening of interactions in dense systems. As such, for vanishingly small $\chi$, guest polymers are expected to disperse as random coils within the coacervate phase, since the coacervate is essentially a good/theta solvent for the guests from the classical polymer physics\cite{de1979scaling, doi1988theory, rubinstein2003polymer, xu2021coil}. 

However, many experiments reveal a surprisingly different reality: Many guest polymers --- including neutral polymers and charged polyelectrolytes/proteins --- experience strong exclusion from the coacervates, even when sharing similar chemical compatibility with the host polymers \cite{marianelli2018impact, mccall2018partitioning, lu2020multiphase, sathyavageeswaran2024self}. The strong guest exclusion means, the guest polymers are effectively attractive to each other within the coacervates phase, and they will from aggregates within the coacervates if no supernatant phase is at present. To counteract the repulsion from the coacervates (effective guest-guest attraction in coacervates) for better encapsulation efficiency, additional charge residues are often required on guest macromolecules \cite{black2014protein, obermeyer2016complex}. Such a strong  tendency for guest aggregations seems to be unique in coacervate systems, since it is not observed in either neutral polymer solutions or in simple polyelectrolyte solutions \cite{wang201750th, Muthukumar2017i50thSolutions}. Given that guest polymers often have comparable size with the coacervate polymers, the mechanism underlying this seemly universal guest aggregations is apparently different from the well-known hydrophobic effect \cite{chandler2005interfaces} or the classical depletion effect \cite{asakura1954interaction,asakura1958interaction}, as both require huge mismatches in length scales between solute and solvent. These observations suggest the existence of a previously unrecognized, strong driving force for guest aggregation within coacervates.

In this study, using extensive molecular dynamics (MD) simulations, we identify a universal mechanism driving this paradoxical aggregation: electrostatic depletion, a strong force arising from electrostatic correlations within the host polyelectrolyte network. Our results demonstrate that, in the absence of chemical incompatibility ($\chi =0 $ across all species), neutral guest polymers spontaneously aggregate within coacervates when their chain length exceeds a few ``electrostatic blobs". Furthermore, the strong electrostatic depletion can drive aggregations of guest polyelectrolytes with low to intermediate charge densities, and fully-charged intrinsically disordered proteins (IDPs), even when these guests energetically favor associations with host polymers. Our findings underscore the significance of electrostatic depletion as a powerful mechanism for controlling macromolecular aggregation and encapsulation in coacervates, with profound implications for the design of synthetic protocells and drug delivery systems.

\section*{RESULTS}
\subsection*{Universal polymer aggregation within complex coacervates}

The complex coacervates in our simulations result from liquid-liquid phase separation of oppositely charged polyelectrolyte solutions (see Fig. S1 in Supplementary Materials), which represent generalized electrostatically-driven coacervate systems in both synthetic and biological contexts (Fig. 1A). We then introduce approximately $1\%$ guest polymers (w/w) to the coacervate matrix to monitor their behavior. To highlight the universality of the electrostatic depletion effect, we set the excluded volume interactions to be the same among all species (polycations, polyanions, water, counterions, and guest polymers), ensuring no chemical incompatibility ($\chi =0 $) between guest and host. 

\begin{figure} 
	\centering
	\includegraphics[width=0.9\textwidth]{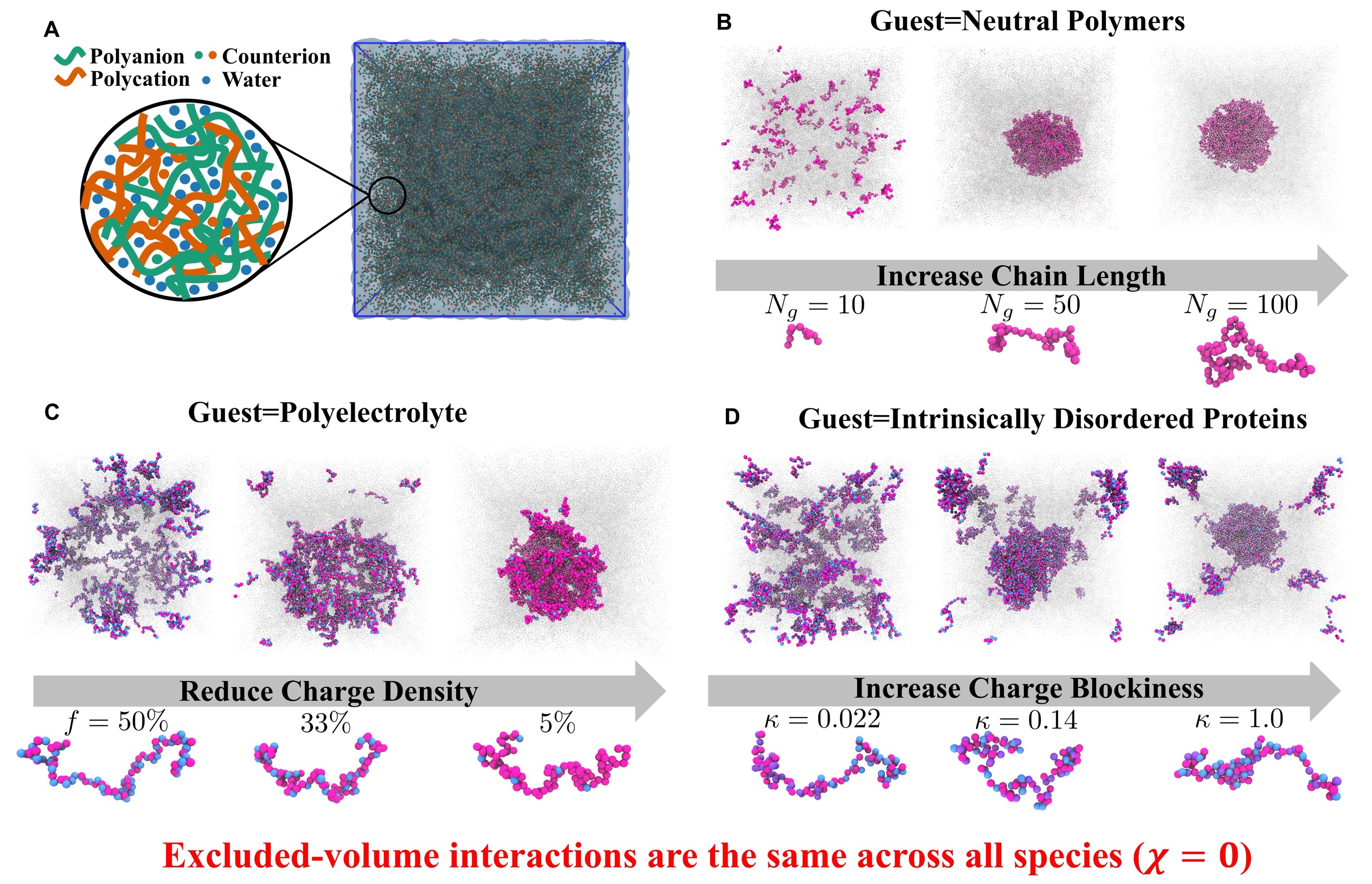}  
        \caption{\textbf{Universal aggregation of neutral polymers, polyelectrolytes, and IDPs within the coacervate phase (without the coexisted supernatant), where $\chi=0$ across all species}. \textbf{A}, The composition of a typical coacervate system (left: schematic; right: simulation snapshot). \textbf{B}, Aggregation/dispersion states of $1\%$ guest neutral polymer with chain length at $N_g=10, 50, 100$.  \textbf{C}, Guest polyelectrolytes of different charge densities ($f= 5\%, 33\%, 50\%$) within the coacervate matrix. \textbf{D}, The behavior of fully-charged IDPs with different charge sequences ($\kappa= 0.0009, 0.014, 1.0$) in the coacervate matrix. Green, orange, and blue (transparent) beads in Figure A represent polycation monomers, polyanion monomers, and water beads, respectively. In Figure B-D, gray beads represent monomers of the background coacervate matrix; magenta, blue, and purple represent the neutral, negatively charged, and positively charged monomers of the guests, respectively. Water and counterions are not shown for clarity. The excluded volume interactions are the same across all species. }
	\label{fig:1}
\end{figure}

Contrary to the classical prediction of mixed chains, we find guest polymers show universal aggregations within the coacervate matrix, particularly pronounced for neutral polymers. When guest chain length exceeds a critical value ($N_g \geq 30$), all neutral chains will immediately aggregate into a globule within the coacervates (Fig. 1B), while only very short chains (e.g., $N_g=10$) remain dispersed (left panel in Fig. 1B). The aggregation trend of guest polymers seems to be independent of the chain length ($N_h$) of host polyions, as the guest polymers ($N_g=50$) show qualitatively similar behavior for both $N_g > N_h=30$ (Fig. 1B) and $N_g < N_h=100$ (Fig. S2 in the Supplementary Materials). Clearly, the underlying mechanism for such aggregation is different from the classical depletion effect \cite{asakura1954interaction,asakura1958interaction}, which requires $N_g \gg N_h$. 

 
More unexpectedly, the ubiquitous aggregation within coacervates also happens to charged guests, including polyelectrolytes carrying same-sign charges and intrinsically disordered proteins (IDP) carrying opposite charges with overall charge neutrality. For guest polyelectrolytes, strong aggregation appears for low-charge-density chains, and the aggregation trend weakens with increasing charge density (Fig. 1C). For instance, at a low charge density of $f=5\%$, all guest polyelectrolytes aggregate into a compact globule, similar to the case for the neutral guest polymers. These polyelectrolytes tend to aggregate until their charge density increases to  $f \approx 50\%$ (where one in every two monomers carries $+1e$ charge, Fig. 1C). Thus, even polyelectrolytes with intermediate charge density will experience a strong ``aggregation force" that clusters them within the coacervate matrix. For IDPs, Fig. 1D  illustrates that aggregation occurs when their charge sequences are less blocky, characterized by the parameter $\kappa$ where a smaller $\kappa$ indicates the charges are more locally balanced, and vice versa \cite{das2013conformations} (see details in Supplementary Materials). For example, IDPs with alternating charges along the chain ($\kappa=0.0009$) aggregate robustly in coacervates, while IDPs with a long-block charge sequence ($\kappa=1.0$) remain dispersed (Fig. 1D). In real IDPs, charges are more sparse (lower charged density), and long-block same charge sequences are less common, therefore we expect more prominent IDP aggregations within coacervates. Indeed, if IDP charge density reduces to $f=50\%$, we find the critical charge blockiness is $\kappa^c \approx 0.014$, below which the IDPs remain aggregated (details in Supplementary Materials). 

Our simulations focus on guest behavior within the single coacervate phase (without the coexisted supernatant phase). Since the excluded volume interactions are the same among all the species, water is a good solvent for the guest polymers. As such, if the supernatant phase is coexisted with the coacervate phases, guest polymers that show aggregations in Fig. 1 will leave the coacervate phase and disperse into the supernatant phase, leading to vanishingly small encapsulation efficiency. Nevertheless, the strong tendency for guest polymers to aggregate within coacervates, even when they are partially charged (polyelectrolytes) or carry both positive and negative charges (IDPs), clearly indicates an inherent, strong driving force in the coacervate that effectively draws guests together. Below, we compute the effective interaction (free energy) between two guest chains within coacervate phase.

\subsection*{Effective guest-guest interaction within coacervates}

\begin{figure} 
	\centering
	\includegraphics[width=0.9\textwidth]{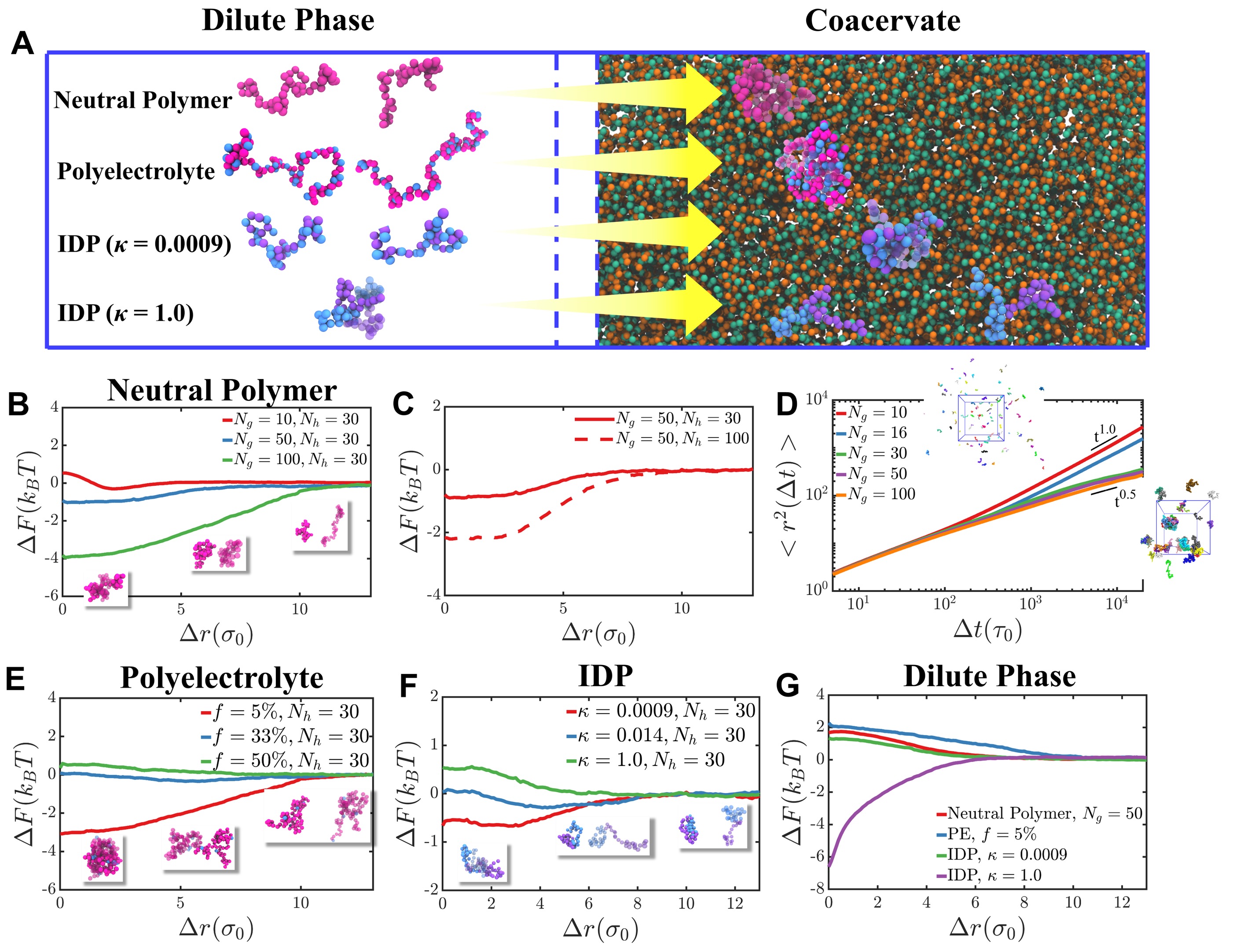} 
        \caption{\textbf{Effective guest-guest interaction within the coacervate phase.} \textbf{A}, Schematic comparisons of guest behavior within the dilute phase (left) and coacervate (right). \textbf{B}, Potential-of-mean-force (PMF) between two neutral guest polymers in coacervate ($N_h=30$) with different guest chain length at $N_g=10, 50, 100$. \textbf{C}, PMFs of two neutral guest ($N_g=50$) in coacervates with different host chain length $N_h=30$ and $N_h=100$.  \textbf{D}, Mean-squared-displacement (MSD) of neutral guest polymers for different $N_g$. The upper and lower insets are representative snapshots for systems with $N_g=10$ and $N_g=100$, respectively, where $N_g=100$ shows clustering of guest chains and exhibit substantial sub-diffusion. \textbf{E}, PMFs of guest polyelectrolytes at charged density $f=5\%$, $33\%$, and $50\%$ in coacervate with $N_h=30$. \textbf{F},  PMFs of fully charged IDPs with different charge sequences  at $\kappa=0.0009$, $\kappa=0.014$, and $\kappa=1.0$. \textbf{G}, PMFs of guest polymers at dilute phase.  Magenta, blue, and purple beads represent the neutral, negatively charged, and positively charged monomers of guest polymers, respectively. The green and orange beads represent the polycations and polyanions in the host coacervates. The insets in the PMF plot are the simulation snapshots at representative separation distances $\Delta r$. }
	\label{fig:2}
\end{figure}

To quantitatively measure the aggregation driving force, we calculate the potential of mean force (PMF) $\Delta F (r)$ between two guest polymers as a function of their center-of-mass distance $\Delta r$ within the coacervate phase. For comparison, we also measure their PMFs in the supernatant phase that coexists with the coacervate phase. Within the coacervate phase, we find that neutral guest polymers with longer chain lengths $N_g$ experience stronger aggregation, evidenced by the deeper and wider free energy wells (Fig. 2B). While very short guest chains ($N_g=10$) exhibit weak repulsion, longer chains become increasingly attractive, with the PMF minimum dropping from $-1.0 k_BT$ for $N_g=50$ to $-4.0 k_BT$ for $N_g=100$, where $k_B$ is Boltzmann constant, and $T$ is temperature.  Consequently, guest chains rapidly aggregate into clusters for $N_g \geq 30$, as manifested by the long sub-diffusion regime in the mean-squared-displacement $\langle r^2 (\Delta t) \rangle$ (Fig. 2D). The attraction strength between the two neutral guest polymers also increases with the chain length of host polyelectrolytes. For a guest polymer with chain length $N_g=50$, their PMF minimum deepens from $-1.0 k_BT$ to $-2.2 k_BT$ when the host chain length increases from $N_h=30$ to $N_h=100$ (Fig. 2C). The effective two-chain attraction gives rise to the universal aggregation of the guest chains, as shown in Fig. 1B.  

The interaction between two low-charge-density ($f=5\%$) polyelectrolytes also shows an attractive PMF, with free energy minimum about ($-3k_BT$), comparable to that of neutral polymers (Fig. 2E). Notably, The inter-chain attraction persists even the charge density increases to $f=33\%$, and only diminish until $f=50\%$. The attractive PMFs also prevail for interactions between fully charged IDPs with low charge blockiness $\kappa$, leading to IDP aggregations within coacervates (Fig. 2A and 2F). For example, IDP with $\kappa=0.0009$ shows an attractive free energy well with a depth of $-0.7k_BT$ (Fig. 2F), and the attractive well becomes shallower as charge blockiness increases from $\kappa=0.0009$ to $\kappa=0.014$, and gradually disappears for $\kappa > 0.014$. When the $\kappa$ increases to $1.0$, the inter-chain interaction becomes repulsive (Fig. 2F), resulting in IDPs dispersing within the coacervate phase (Fig. 2A).  

Interestingly, guest polymers exhibit drastically different behavior in the coacervate phase compared to the dilute phase (Fig. 2A). In good solvent conditions (as in this work), both neutral polymers and low-charge-density polyelectrolytes are repulsive in the dilute phase (Fig. 2G), leading to dispersed configurations (left panel in Fig. 2A and Fig. S3 in Supplementary Materials). In contrast, both types aggregate as a compact globule in the coacervate phase. This opposite trend also applies to IDPs: In the dilute phase, it is known that IDPs with high $\kappa$ such as $\kappa=1.0$ will undergo self-coacervation \cite{das2013conformations, nott2015phase, madinya2020sequence} to form aggregates, as evidenced by their attractive PMF (Fig 2G). Conversely, these IDPs become repulsive within the coacervate phase (Fig. 2A and 2F). On the other hand, low-$\kappa$ IDPs are repulsive within dilute phase but become attractive in the coacervate phase (Fig. 2F and 2G). Clearly, the densely charged, crowded coacervate environment significantly alters both the structure and thermodynamics of guest macromolecules.

\subsection*{Origin of the electrostatic depletion}

The electrostatic depletion effect arises from the unique microscopic charge-charge correlation within the complex coacervates, where the host polyelectrolytes can be described by a sequence of electrostatic ``blobs" of size $\xi$, with each blob preferentially surrounded by its oppositely charged counterpart\cite{shusharina2005scaling,rumyantsev2018complex,rubinstein2018structure} (Fig. 4A). This structural correlation between oppositely charged electrostatic blobs has been confirmed by recent scattering experiments and theory \cite{fang2023scattering}. Although the blob concept should only be rigorously valid for weakly charged systems \cite{rumyantsev2025electrostatically}, the qualitative picture apply to our systems in explaining the origin of electrostatic depletion. Here, we demonstrate how such charge correlation gives rise to electrostatic depletion that promotes guest aggregation. For simplicity, we examine the electrostatic depletion effect when the guests are neutral polymers.

\begin{figure} 
	\centering
    	\includegraphics[width=0.95\textwidth]{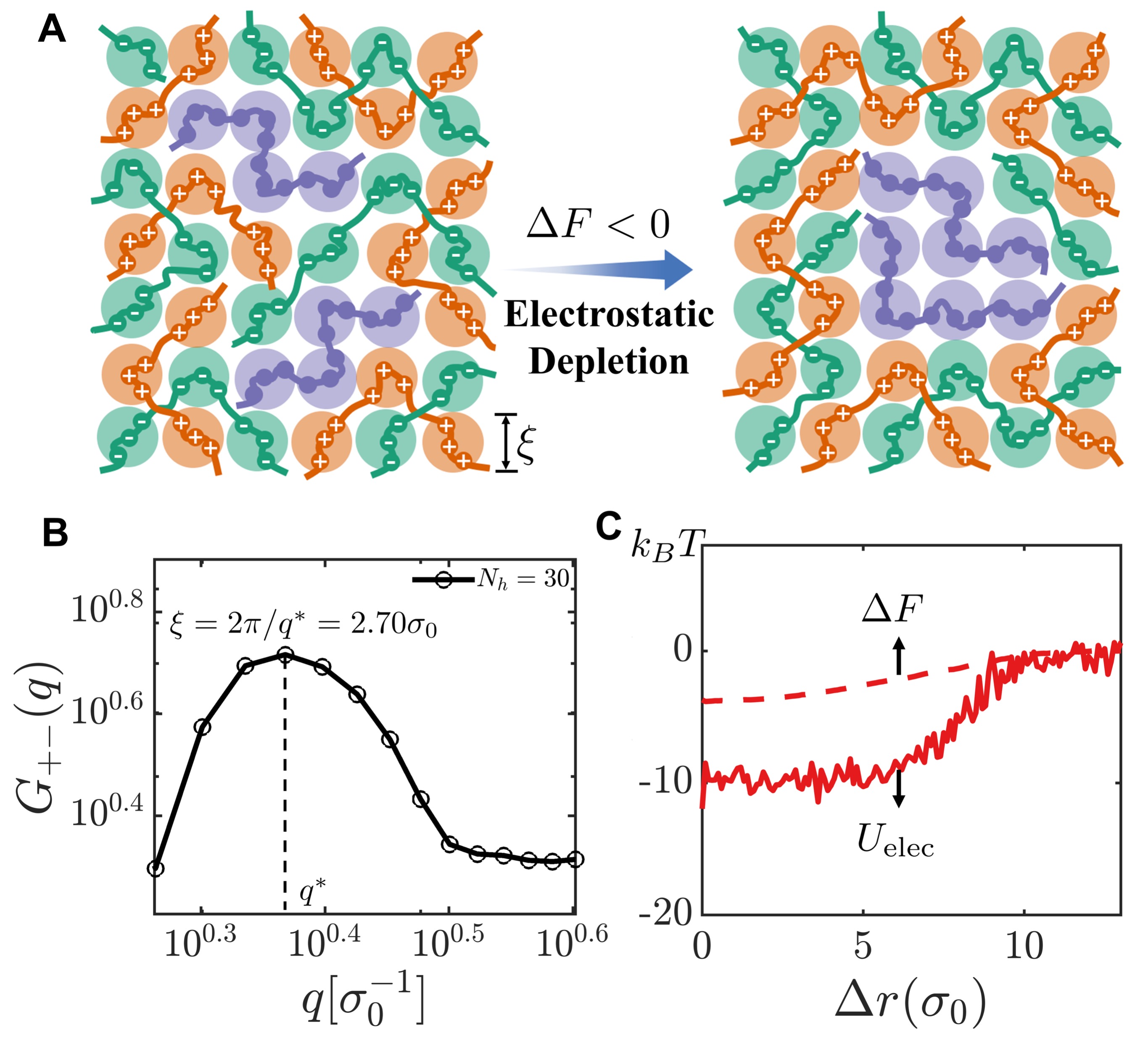} %
            \caption{\textbf{Origin of the electrostatic depletion}. \textbf{A}, Schematic illustrations of electrostatic correlation of oppositely charged blobs with size $\xi$ within the coacervate matrix give rise to favorable free energy changes ($\Delta F$) upon guest aggregation. Polyanion blobs (green) are mostly surrounded by polycation blobs (orange) due to the existence of charge correlations, where guest polymers (purple blobs) can create interfaces with the charged blobs. \textbf{B}, Electrostatic blob length $\xi = 2.7\sigma_0$ calculated from the Fast Fourier Transform (FFT) of the radial distribution functions between the polycations and polyanions. \textbf{C},  PMF ($\Delta F$, dashed line) and electrostatic energy change ($U_{\text{elec}}$, solid line) for two neutral guest chains ($N_g=100$) interacting within a coacervate matrix ($N_h=30$).}
	\label{fig:4_1}
\end{figure}

In the absence of guests, the interaction (free) energy between two neighbored electrostatic blobs equals to thermal energy $-k_BT$, according to scaling theory \cite{shusharina2005scaling,rubinstein2018structure}. Introducing neutral guest polymers into the coacervate matrix inevitably creates interfaces that separate the close associations of neighboring oppositely-charged blobs (Fig. 4A), with surface tension $\gamma \sim k_BT/\xi^2$ (here we adopt the convention of scaling theory in polymer physics and omit the prefactor). In good solvent conditions, the guest polymers can be represented by a sequence of concentration blobs with the same size of electrostatic blob \cite{wang2006regimes,rumyantsev2018complex}. Our simulation shows the blob size is approximately $\xi=2.7\sigma_0$ (Fig. 4B, also see in Supplementary Materials). Considering $n$ neutral guest polymer chains initially disperse as random coils within a coacervate matrix (left panel in Fig. 4A), each chain will have contour length $\frac{N_g}{g} \xi$, with $g$ being the number of monomers in a blob satisfying $\xi \sim g^{3/5}$. The total interfacial energy between coacervate matrix and the dispersed guest polymers is $E_d = \gamma \cdot n \cdot \pi \xi \cdot \frac{N_g}{g} \xi = \gamma n \pi \frac{N_g}{g} \xi^2$. When all guest polymers aggregate into one single globule, the interfacial energy can be easily calculated as $E_a= \gamma (n\frac{N_g}{g})^{2/3} \pi\xi^2$. Consequently, the electrostatic depletion energy can be expressed as the difference in interfacial energy between the dispersed and aggregated states: 
\begin{equation}
    \Delta F = E_a -  E_d = \gamma \pi \xi^2 \left ( (n\frac{N_g}{g})^{2/3} - n\frac{N_g}{g} \right) \sim k_BT \left ( (n\frac{N_g}{g} )^{2/3} - n\frac{N_g}{g} \right) \label{eq1}
\end{equation} for $N_g \gg g $. 

Equation (1) shows $\Delta F < 0$, favoring aggregation. This remains valid as long as the size of guest polymers in the coil state spans over a few electrostatic blobs (allowing the guest to create separation between neighboring electrostatic blobs). As such, except for very short chains (e.g., $N_g=10$), there is always a significant free energy drop ($\Delta F \ll -k_BT$) promoting guest aggregation, with the depletion strength increasing with guest chain length, consistent with our findings in Fig. 1B and Fig. 2B. Equation (1) also indicates that the electrostatic depletion is stronger for higher host polymer concentration, as larger host concentration results in smaller $g$ \cite{de1979scaling, doi1988theory, rubinstein2003polymer}. Indeed, our simulation shows that the guest attraction is stronger in denser coacervate ($N_h=100$, polymer fraction $c_p=38\%$) than in looser coacervates ($N_h=30, c_p=33\%$), as shown in Fig. 2C and Fig. S1. 
            
To further confirm the origin of electrostatic depletion, we calculate the change of the electrostatic interaction ($\Delta U_\text{elec}$) alongside the guest-guest PMF in coacervates ($N_g=100, N_h=30$). Figure 4C shows $U_\text{elec}$ significant drops when the two neutral polymers come together, which is the main reason for the decrease of free energy $\Delta F$, as other interaction potentials (excluded volume and bond interactions, see Fig S5 in Supplementary Materials) remain nearly unchanged. We also note that $\Delta U_\text{elec} \approx -10 k_BT$ drops deeper than the drop in free energy $\Delta F \approx -4k_BT$, indicating this two-chain complexation associates with a penalty in configuration entropy ($T\Delta S \approx -6k_BT < 0$). This entropy penalty is expected for two-chain systems, as the guest chains transition from a coil state to the collapsed state upon complexation; this penalty will diminish for many-chain aggregations as the chains recovers to ideal coil state in the larger aggregate,  making equation (1) applicable. The significant drop in $\Delta U_\text{elec}$ emphasizes the electrostatic origin of this depletion force, as indicated by its name. The impact of electrostatic depletion on single guest chains are given in Supplementary Materials \cite{supplementalmaterial}.       

\subsection*{Implications of the electrostatic depletion to the guest encapsulation efficiency}

\begin{figure} 
	\centering
    	\includegraphics[width=0.95\textwidth]{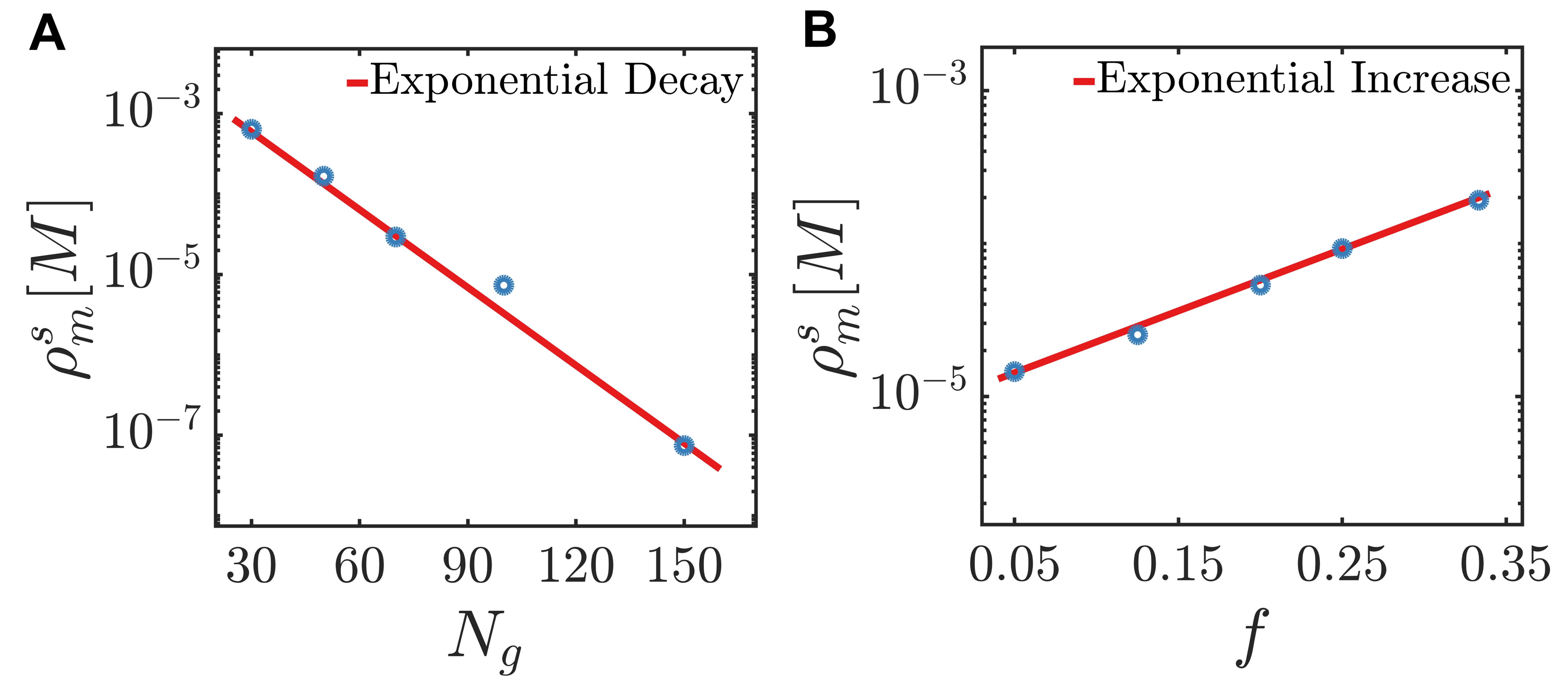} %
            \caption{\textbf{The consequence of electrostatic depletion on the guest encapsulation efficiency}. \textbf{A}, Spinodal concentration $\rho_m^s$ of neutral guest polymers in coacervate as function of their chain length $N_g = 30 \sim 150$. \textbf{B}, Spinodal concentration $\rho_m^s$ of polyelectrolytes with $N_g=100$ in coacervate as a function of their charge density $f = 5\%\sim33\%$}
	\label{fig:4_2}
\end{figure}

The strong electrostatic depletion has direct implications for the guest encapsulation efficiency of coacervates, which is crucial for many applications relying on the enrichment of guest molecules. Here, we define guest encapsulation efficiency as the onset monomer concentration for spontaneous phase separation of guest polymers, beyond which the guest polymers can no longer remain dispersed within the coacervate phase. Thermodynamically, this concentration is the spinodal concentration $\rho_m^s$. Taking the osmotic pressure of guest polymers as virial expansion of guest density, the spinodal concentration $\rho_m^s$ relates to the second osmotic virial coefficient $B_2$ as $\rho_m^s=-\frac{1}{2B_2}$, where $B_2$ can be calculated via guest PMF as \cite{wang201750th,chen2022driving}:
\begin{equation}
B_2=-2\pi \int \left( e^{-\Delta F(r)/k_BT} -1\right)r^2dr
\end{equation}

Figure 4D shows  $\rho_m^s$ decreases almost exponentially with increasing guest chain length $N_g$. As such, while smaller molecules $N_g < 10$ can be easily enriched within coacervate phase, the encapsulation efficiency drastically decreases for longer chains. For any guest polymers longer than $N_g = 50$, $\rho_m^s = 1.67\times 10^{-4} M$ --- a vanishingly small encapsulation efficiency for practical applications. On the other hand, in the case of guest polyelectrolytes, its spinodal $\rho_m^s$ in coacervates increases exponentially with its charge fraction $f$ (Fig. \ref{fig:4_2}E), suggesting that the encapsulation efficiency can be substantially improved by grafting more charged residues, consistent with experiments \cite{blocher2017complex, mctigue2019design, agrawal2025charge}.

\section*{DISCUSSION}
This study demonstrates that electrostatic depletion --- an intrinsic force arises from charge-charge correlation within coacervates --- can robustly drive aggregations of guest polymers,  even in the absence of chemical incomparability. The electrostatic depletion lead to effective attractions between the guest chains (repulsion from the coacervate phase), with the strength increasing with guest chain length. As such, while coacervates are known for its efficiency in encapsulating small molecules \cite{johnson2014coacervate,lv2015photocatalytic,anraku2016systemically,egbeyemi2024transforming}, encapsulating macromolecules becomes challenging without adding more charge residues.    

At first glance, the thermodynamic origin of electrostatic depletion seems to be energy-driven, as the aggregation accompanies with significant drops in electrostatic interaction potential $\Delta U_\text{elec}$. This is different from the traditional depletion effect (colloidal aggregations in polymer solutions), which is entropy-driven \cite{asakura1954interaction, asakura1958interaction}. However, by exploiting the temperature-dependence of water dielectric constant, recent work by Chen and Wang show that electrostatic interactions in aqueous environments have an entropy origin from water reorganization \cite{chen2022driving}. Specially, the entropy component in electrostatic interaction is $T\Delta S_{elec}= -1.36\Delta U_\text{elec}$, which completely dominates the total free energy changes. This ``electrostatic entropy" has been shown to be a major entropic driving force for polyelectrolyte complex coacervation. By this argument, electrostatic depletion should be considered entropy-driven. On the other hand, coacervates are densely populated with $25 \% \sim 45\%$ hydrated polyelectrolytes. Therefore, it is unclear whether the electrostatic entropy (based on continuum electrostatics in water) can still apply to these dense systems. The thermodynamic origin (entropy v.s. energy) of electrostatic depletion warrants further investigation. 

The electrostatic depletion mimics a larger-scale ``solvation" effect within coacervates, where the structural correlation among electrostatic blobs acts as an ``associative bonding network", analogous to hydrogen bonding network in water, and disruption of this network by introducing guests will result in solute aggregation. The discovery of electrostatic depletion fills the gap in the aggregation mechanisms between nanoscale hydrophobic effect (water molecule level) and the macroscopic depletion effect (mesoscale particle level). Unlike the traditional depletion effect that requires mismatched length scales, electrostatic depletion is relevant in fluid systems where solute and solvent are both polymers with comparable sizes. Furthermore, our findings provide insights for the rational design of coacervate systems essential for synthetic and biological applications. For instance, we can fine-tune the charge density and sequence between guest and host polymers to ensure that  ($N_g \leq g$), thereby avoiding electrostatic depletion and enabling high encapsulation efficiency.

\section*{Acknowledgments}
Z. Wu and S. Chen thank for the start-up funding from The Hong Kong University of Science and Technology (HKUST). We also thank for the general allocation of the computational time from the HPC4 cluster in the Information Technology Services Center (ITSC) of HKUST. 


\paragraph*{Author contributions:}
\textbf{Zongpei Wu} (simulation $\&$ theory): Conceptualization, Methodology, Software, Investigation, Visualization, Writing - original draft. contributed to the data analysis and ran the simulations. \textbf{Shensheng Chen}: Conceptualization, Visualization, Supervision, Writing – review $\&$ editing, Funding acquisition.

\paragraph*{Data and materials availability:}
All data is included in the article and/or in the Supplementary Materials \cite{supplementalmaterial}.

\subsection*{Supplementary materials}
Supplementary sections S1 to S8\\
Figs. S1 to S6\\
References [18-21, 30, 50-53]



\bibliography{nature_preprint}

\bibliographystyle{Science}


\newpage


\renewcommand{\thefigure}{S\arabic{figure}}
\renewcommand{\thetable}{S\arabic{table}}
\renewcommand{\theequation}{S\arabic{equation}}
\renewcommand{\thepage}{S\arabic{page}}
\setcounter{figure}{0}
\setcounter{table}{0}
\setcounter{equation}{0}
\setcounter{page}{1}


\begin{center}
\section*{Supplementary Materials for\\ \naturetitle}
Zongpei Wu,$^{1}$ Shensheng Chen$^{1\ast}$\\
\normalsize{$^{1}$Department of Chemical and Biological Engineering,} \\
\normalsize{The Hong Kong University of Science and Technology,} \\
\normalsize{Clear Water Bay, Kowloon, Hong Kong, China}\\
\normalsize{$^\ast$Corresponding authors. E-mail: shensheng@ust.hk} \\
\end{center}

\subsubsection*{This PDF file includes:}
Supplementary sections S1 to S8\\
Figs. S1 to S6\\
References

\newpage


\subsection*{S1 Simulation Details} 

In this study, we employ the Gaussian Core Model with Smeared Electrostatics (GCMe) \cite{ye2024gcme} to perform coarse-grained simulations to investigate coacervate systems. Following the default setup of the GCMe model, we set the reduced number density to $\rho=2.5/\sigma_0^3$, where $\sigma_0 \approx 0.64$nm is the size of coarse-grained beads for all the species (host polyelectrolyte monomers, guest chain monomers, counterions, and solvent molecules) \cite{ye2024gcme}. The excluded volume interaction between bead $i$ and $j$ separated by distance $r_{ij}$ is modeled by a Gaussian soft-core repulsive potential:
\begin{equation}
    U_{\text{ex}} \left ( r_{ij}\right )=A_{ij}\left ( \frac{3}{2 \pi \sigma_{ij}^2}\right )^{3/2} \exp \left( -\frac{3r_{ij}^2}{2\sigma_{ij}^2} \right )
\end{equation} 
Here, we set $A_{ij}=25.2k_BT \sigma_0^3$ the same for all beads, where $k_BT$ is the thermal energy at room temperature, $\sigma_i=\sigma_j=0.5\sigma_0$ is the mass smearing radii of coarse-grained beads, and  $\sigma_{ij}=\sqrt{\sigma_i^2 + \sigma_j^2}=0.71\sigma_0$. Since $A_{ij}$ is the same for all the beads, there is no differences in hydrophobicity across species.   

Long-range electrostatic interaction between two charged beads follows a modified, gaussian-core Coulomb potential:
\begin{equation}
    U_\text{elec} \left ( r_{ij}\right )= \frac{e^2}{4 \pi \epsilon_0 \epsilon_r r_{ij}} \text{erf} \left( \frac{\pi^{1/2}r_{ij}}{2^{1/2}a_{ij}} \right ),
\end{equation}
where $a_{ij}=\sqrt{a_i^2+a_j^2}=0.71\sigma_0$, $a_i=a_j=0.5\sigma_0$ is the charge smearing radii, $\epsilon_0$ is the vacuum permittivity, and $\epsilon_r$ is the relative permittivity of water at room temperature. The corresponding Bjerrum length $l_B= \frac{e^2}{4\pi\epsilon_0\epsilon k_BT} \approx 1.1\sigma_0$ is set to match the dielectric environment of water at room temperature. 

Neighboring polymer monomers along a single chain (including polyion and guest polymer chain) are connected via a harmonic bond potential:
\begin{equation}
E_{\text{bond}}=\frac{1}{2}K_\text{bond}(r-b)^2,
\end{equation}
where $K_\text{bond}=100k_BT/\sigma_0^2$ is the bond strength. For fully charged polyelectrolytes, the electrostatic strength is $l_B/b \approx 1.0$, falling into intermediately charge regime.  


All simulations are performed in a simulation box with a dimension $50\sigma_0 \times 50\sigma_0 \times 50\sigma_0$, using the LAMMPS \cite{plimpton1995fast} platform. The characteristic time scale is given by $\tau_0=\sqrt{m_0\sigma_0^2/(k_BT)}=7.8$ps, where $m_0 \approx 72$g/mol to represent to coarse-grained level of one water bead contains 4 water molecules. The integration time step is set to $\delta t=0.01 \tau_0$. 

\subsection*{S2 Prepare Complex Coacervates in Simulation} 

We obtain the coacervates compositions from liquid-liquid phase separation (LLPS) of polycation and polyanion mixture in a $120\sigma_0 \times 30\sigma_0 \times 30\sigma_0$ simulation box (Fig. S1). In this study, we investigate coacervate matrix with host chain lengths of $N_h=30$ and $N_h=100$. For the coacervate with host chain length $N_h=30$, the LLPS system contains $n_c=n_a=500$ fully charged polycations and polyanions, with $2700$ small ions and the remaining water beads initially disperse within the simulation box. For the coacervate system with $N_h=100$, we set $n_c=n_a=200$, together with the same number of small ions as before. Both systems are equilibrated for $3\times10^4\tau_0$ with fixed center-of-mass of the polyions at $x=60\sigma_0$ (the snapshots of Fig. S1 A and B).

We quantify the equilibrium composition by computing the density profile along the $x$ direction. Specifically, we divide the simulation box into $120$ bins of thickness $1\sigma_0$ along the $x$ direction and count the polyions, water, and small ions, respectively in the bins. For $N_h=30$, the coacervate contains around $33\%$ polyions, $66\%$ water, and $1\%$ small ions, while coacervate with $N_h=100$ contains $38\%$ polyions, $61\%$ water, and $1\%$ small ions (Fig. S1 A and B). 

Based on the equilibrium profiles from these LLPS simulations, we build bulk coacervate systems in simulation box with dimension $50\sigma_0 \times 50\sigma_0 \times 50\sigma_0$ by replicating the coacervate compositions for $N_h=30$ and $N_h=100$. An example of the resulting coacervate matrix for $N_h=30$ is shown in Fig. 1A in the main text. 

Guest species--- including neutral polymers, locally charged polyelectrolytes, and intrinsically disordered proteins (IDPs) --- initially evenly distributed within the coacervate matrices. In each simulation, we study the guest behavior by putting $0.3\% \sim 1.6\%$ guest polymers (replace the water molecules) within the coacervate matrix, with the assumption that they do not significantly alter the properties of the bulk coacervates. The charged density of the guest polyelectrolytes $f$ is defined as the fraction of the charged monomers, which are evenly distributed along the chain. For example, $f=33\%$ corresponds to one charged monomer per three neighboring monomers along the chain. 

Fully-charged IDPs are prepared by carefully positioning equal number of positive and negative charges along a backbone of 50 monomers. In this work, we use the order parameter $\kappa$ introduced by Das and Pappu \cite{das2013conformations} to characterize the IDP charge sequences . Following the standard procedure, we calculate $\kappa$ by dividing the chain into $N_{\text{s}}$ overlapping segments of size $s$. For each segment, we first calculate its charge asymmetry as $\lambda_i=\frac{(f_+-f_-)_i^2}{(f_++f_-)_i}$, where $f_+$ and $f_-$ is the fraction of positively and negatively charged monomers within segment $i$. We then quantify the squared deviation from perfect charge segregation as $\delta=\frac{\sum_i^{N_{\text{s}}}(\lambda_i-\lambda)^2}{N_{\text{s}}}$, where $\lambda$ is the overall charge asymmetry defined as $\lambda=\frac{(f_+-f_-)^2}{(f_++f_-)}$. Here, we use the deviation from $\delta_{\text{max}}$ ($\delta_{\text{max}}$ corresponding to the sequence where half side of the chain is positively charged and the other side is negatively charged.) to define $\kappa = 1 - \delta/\delta_{\text{max}}$, ensuring $0 \leq \kappa \leq 1$. We calculate $\kappa$ using two segment sizes: $s = 5$ and $s = 6$ and take the average of these two values as the final $\kappa$ for each charged blockiness sequence variant \cite{das2013conformations}. In Figure 1D of the main text, the $\kappa = 1.0$ means fully segregated charges (all positive followed by all negative), where $\kappa =0.0009$ corresponds to alternative positive-negative distribution.  

For any charged guest, additional counterions matching the guest charges are introduced to maintain charge neutrality. For each system, we run $2\times10^5\tau_0$ to reach equilibrium state.

\begin{figure} 
	\centering
	\includegraphics[width=0.9\textwidth]{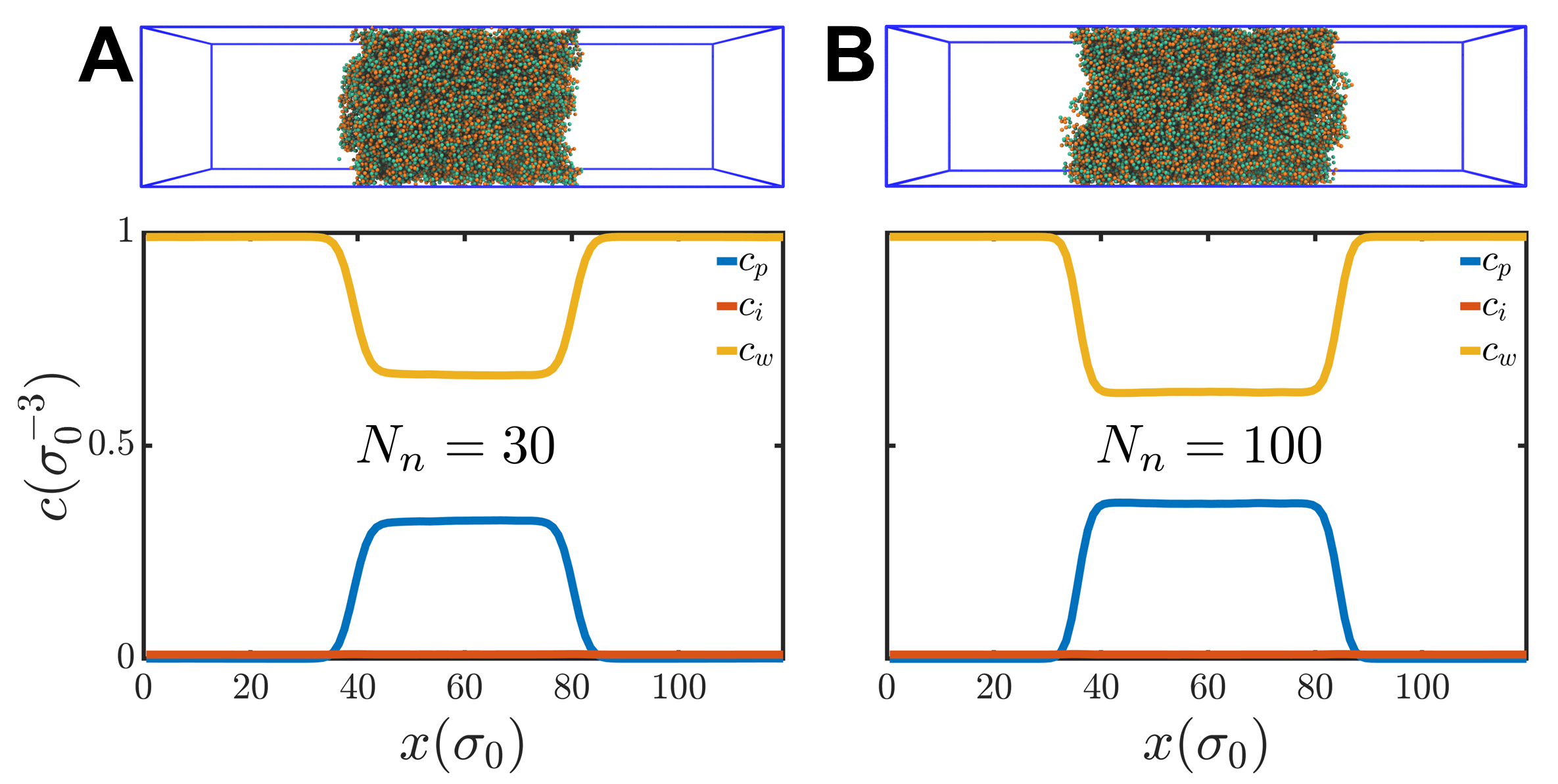} 
        \caption{The composition of the coacervate matrix obtained from liquid-liquid phase separation between oppositely charged polyelectrolytes with symmetric chain length $N_h=30$ (\textbf{A}) and $N_h=100$ (\textbf{B}), respectively. $c_p, c_i, c_w$ represent repetitively the fractions of polymer, small ions, and water. The upper panels are the snapshots, where the green and orange beads represent the polyanions and polycations, respectively. The water molecules and counterions are not shown for clarity.}
	\label{supfig:1}
\end{figure}

\subsection*{S3 PMF Calculation Between Two Guest Polymers}

In this study, the interaction between two guest molecules is quantified by their potential of mean force (PMF) $\Delta F$ as a function of their center-of-mass distance $\Delta r$. We calculate the PMF profiles by using the adaptive bias force (ABF) algorithm \cite{plimpton1995fast, darve2008adaptive} implemented in the LAMMPS platform. The range of $\Delta r$ is $0\sigma_0 \sim 15\sigma_0$. To improve the sampling efficiency and accuracy, this distance range is partitioned into five continuous sampling windows: $0\sigma_0\sim1\sigma_0$, $1\sigma_0\sim3\sigma_0$, $3\sigma_0\sim6\sigma_0$, $6\sigma_0\sim10\sigma_0$, $10\sigma_0\sim15\sigma_0$ \cite{comer2015adaptive}. Each window is further divided into smaller bins with equal width of $0.1\sigma_0$. The PMF in each window has run for $1\times10^6 \tau_0$ to ensure convergence. All the PMF calculations are performed within  a $24\sigma_0 \times 12\sigma_0 \times 12\sigma_0$ simulation box, with two guest chains (and their counterions, if any) in the respective environments (coacervate phase or dilute phase).

\subsection*{S4 Guest Behavior Within Coacervate with $N_h=100$}

Figure S2 shows the neutral polymer behavior in the coacervate matrix of $N_h=100$. Similar to the case of $N_h=30$ detailed in the main text, the guest polymers exhibit apparent aggregation behavior for chain length $N_g \geq 50$. Notably, the neutral polymer can spontaneously aggregate even for $N_g < N_h$. Therefore, the origin of this guest aggregation is not from the classical depletion effect, which requires $N_g \gg N_h$.

\begin{figure} 
	\centering
	\includegraphics[width=0.9\textwidth]{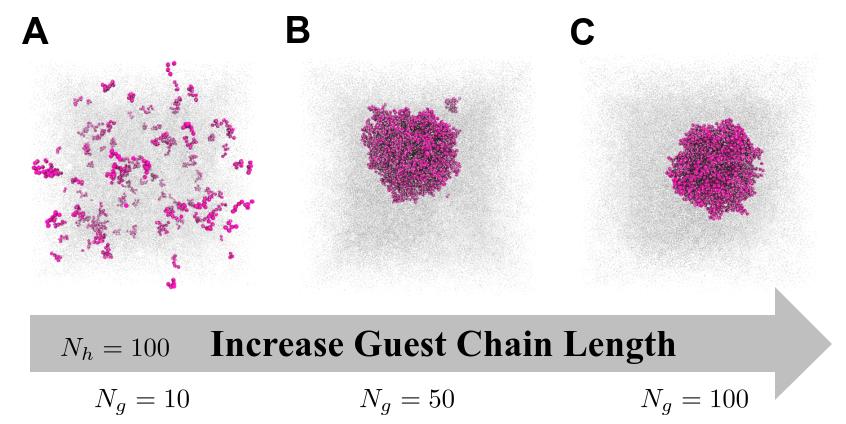} 
        \caption{The behavior of neutral guest polymers with different chain length (\textbf{A}, $N_g=10$; \textbf{B}, $N_g=50$; \textbf{C}, $N_g=100$) within the coacervate matrix with $N_h=100$. Gray and magenta beads represent monomers of the background coacervate matrix and neutral monomers, respectively. The water and counterions are not shown for clarity.}
	\label{supfig:2}
\end{figure}

\subsection*{S5 Guest Behavior Within Dilute Phase}

To highlight the electrostatic depletion is a unique phenomenon in the coacervate phase, we simulate the guest behavior in the dilute phase environment as a comparison. In the dilute phase, both neutral polymers and polyelectrolytes exhibit dispersed distribution due to repulsive intermolecular interactions (Fig, 2G in main texts), as shown in Fig. S3 A and B. This is opposite compared to the cases in the coacervate phase, where the neutral polymers and low-charge-density polyelectrolytes show universal aggregation behavior (Fig. 1 B-C in the main text). For guest IDPs within the dilute phase, high $\kappa$ sequence such as $\kappa=1.0$ firstly self-coacervation to form a small globule, then aggregate to a larger compact globule (Fig. S3 D). However, they become repulsive in the coacervate phase, as detailed in Fig. 2 in the main text. Low $\kappa$ such as $\kappa=0.0009$ are repulsive and remain dispersed within the dilute phase (Fig. S3 C), while they aggregate within the coacervate phase. Clearly, all the guest chains, including neutral polymers, polyelectrolytes, and IDPs, behave drastically different in the coacervate phase compare to that in the dilute phase.

\begin{figure} 
	\centering
	\includegraphics[width=0.9\textwidth]{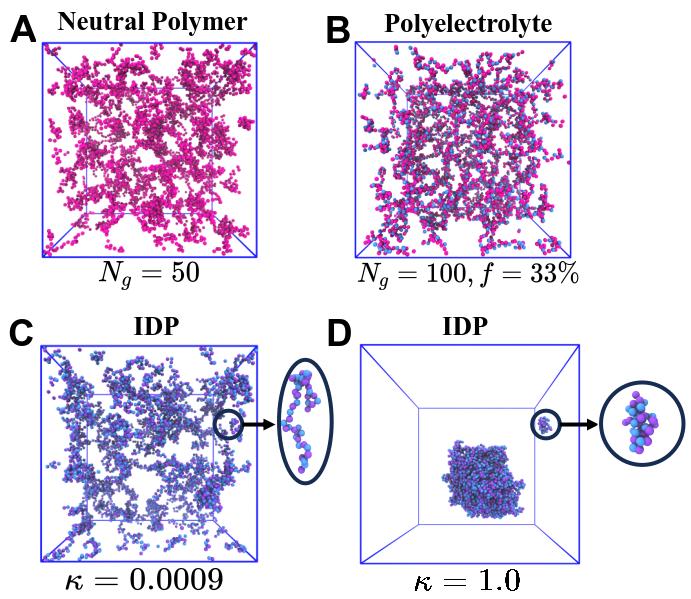} 
        \caption{The behavior of around $1\%$ guest polymers in the dilute phase. \textbf{A}, Neutral polymer with $N_g=50$; \textbf{B}, Polyelectrolyte with $N_g=100$ and $f=33\%$; \textbf{C} and \textbf{D}, IDP with $\kappa=0.0009$ and $\kappa=1.0$, the insets show representative equilibrium configurations of single-chain conformations. Magenta, blue, and purple represent the neutral, negatively charged, and positively charged monomers, respectively. The water are not shown for clarity.}
	\label{supfig:3}
\end{figure}

\subsection*{S6 The Behavior for aggregation of $50\%$ Charged IDP Guest}

While fully charged IDPs with very low $\kappa$ (about $0.014$) remain dispersed within the coacervate matrix, biologically relevant IDPs often exhibit aggregation and compartmentalization. We suppose this is because, in real IDP systems, not every monomer is charged, and we expected more prominent IDP aggregations within the coacervate. To this end, we examine IDPs chain length of $N_g=100$, where half of the monomers carry equal amount (25) of positive and negative charges. 

(Figure. S4 B) shows that for IDPs with charged density $f=50\%$, strong aggregation appears for sequence with $\kappa=0.14$, while the same $\kappa$ in the fully-charged IDPs will exhibit dispersed distribution. Thus, aggregations will be more common for real IDPs whose charged density is generally lower than the ones in the main text.   

\begin{figure} 
	\centering
	\includegraphics[width=0.75\textwidth]{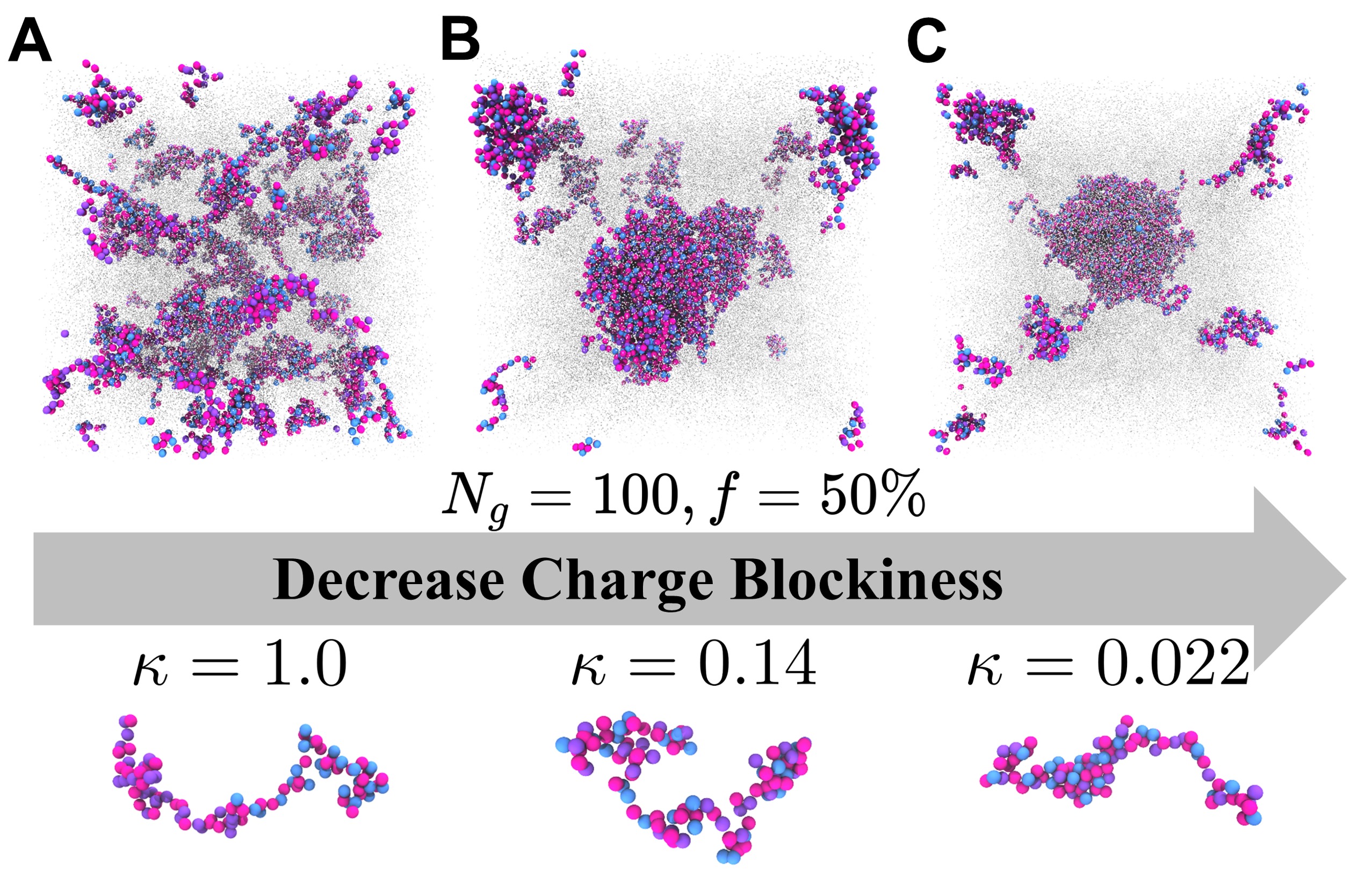}
        \caption{The behavior of $50\%$ charged guest IDPs in the coacervate matrix ($N_h=30$). Gray beads represent monomers of the background coacervate matrix; magenta, blue, and purple represent the neutral, negatively charged, and positively charged monomers, respectively. The water and counterions are not shown for clarity.}
	\label{supfig:4}
\end{figure}




\subsection*{S7 Decomposition of the PMF Profile}

To further demonstrate the electrostatic depletion comes from charge-charge correlation within the coacervate matrix, we decompose the free energy $\Delta F$ into the electrostatic energy, the excluded volume energy $U_\text{ex}$, and polymer bond energy $U_\text{bond}$ contribution. Figure S5 shows the most significant energy change underlying guest aggregation is the electrostatic interaction contribution, while both the excluded volume and bond interactions exhibit negligible variation (oscillating around $0k_BT$).

\begin{figure} 
	\centering
	\includegraphics[width=0.75\textwidth]{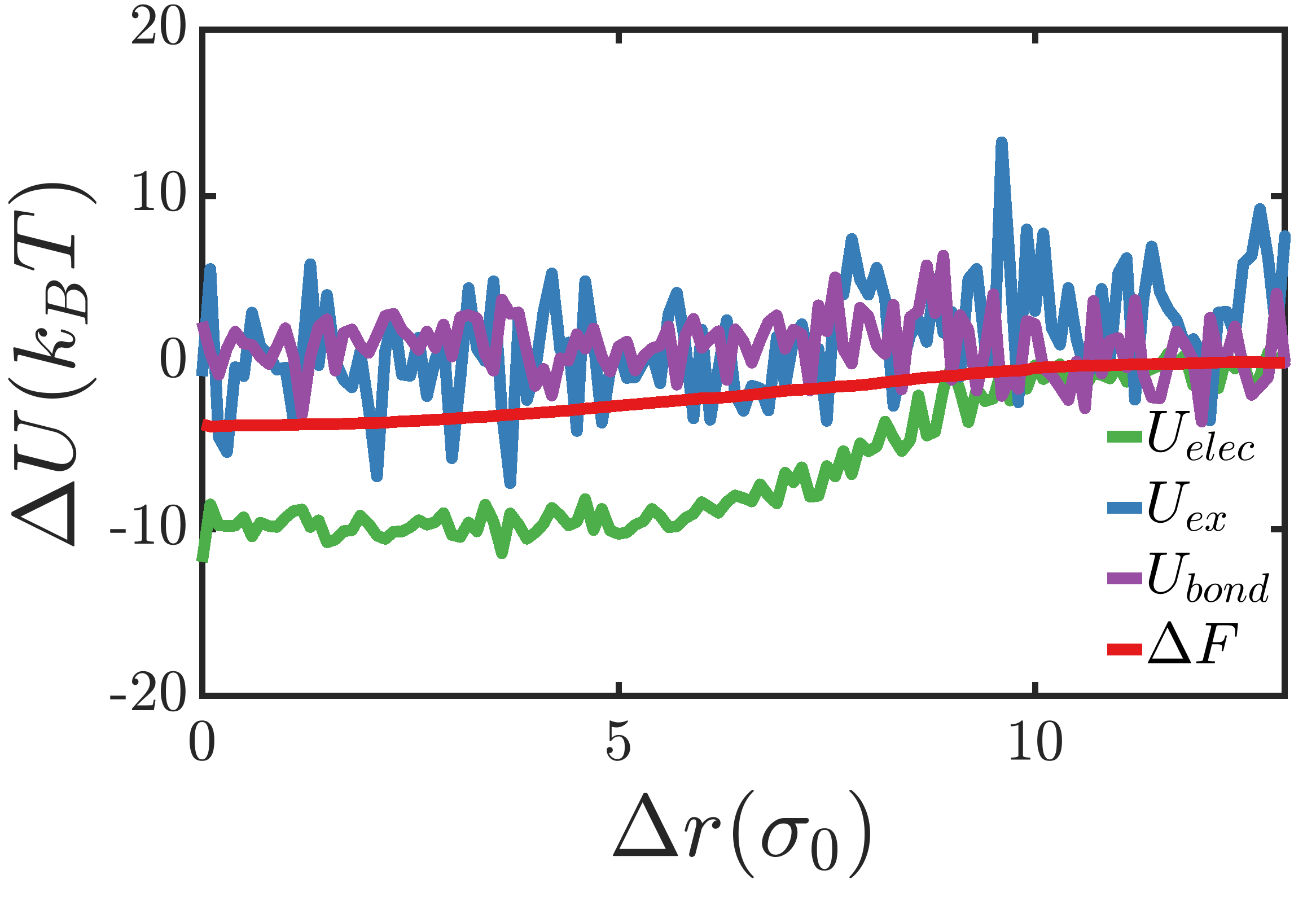} 
        \caption{Decomposition of the energy components alongside PMF of neutral guest polymer ($N_g=100$) within the coacervate matrix ($N_h=30$).}    
	\label{supfig:5}
\end{figure}

\subsection*{S8 Compression of a Single Chain in the Coacervate Phase}

Electrostatic depletion also significantly influences the single chain morphology, which is otherwise predicted to be as random coil in the absence of difference in excluded volume interaction between guest and host polymers \cite{rubinstein2003polymer, de1979scaling, doi1988theory, xu2021coil}. To highlight the electrostatic effect, we calculate the end-to-end distance of a single neutral guest chain, normalized by its ideal chain conformation ($N_g^{1/2}b$). Figure S6 shows that, unless for very short chains ($N_g < 30$), the guest chain is compressed within the coacervate phase, and the compression effect enhances with increasing chain length. 

\begin{figure} 
	\centering
	\includegraphics[width=0.75\textwidth]{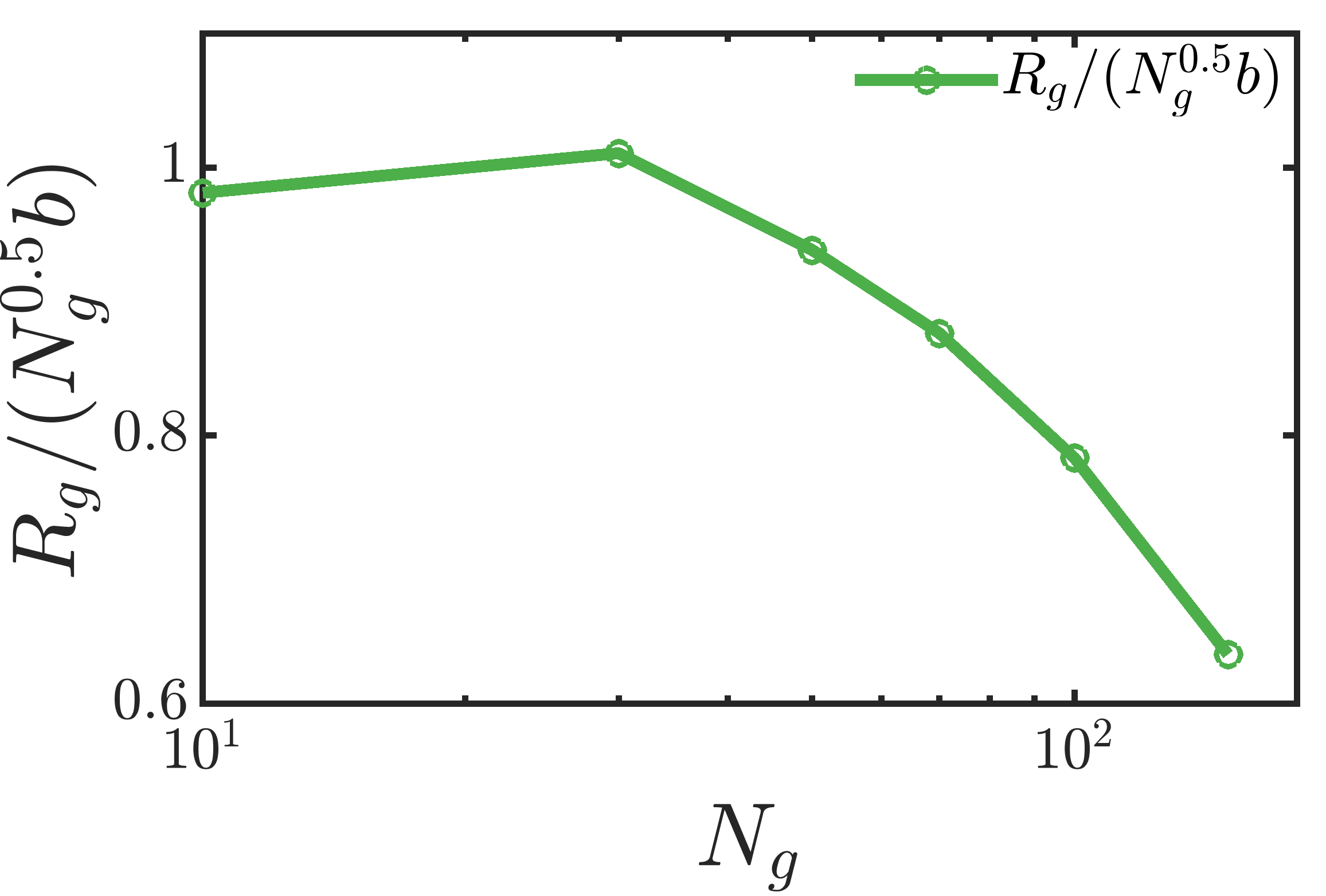} 
        \caption{The end-to-end distance of a single neutral guest chain ($R_g$) with different chain length $N_g$ within the coacervate phase ($N_h=30$). Data are normalized by the ideal chain prediction $N_g^{1/2}b$, where $b$ is the bond length.}   
	\label{supfig:6}
\end{figure}

  
\newpage

\end{document}